# Extraordinary Tunability of the Superexchange Interactions in Nanoparticles of the Multiferroic 0.50BiFeO$_3$-0.50PbTiO$_3$

Chandan Upadhyay[a], Pappu Kumar Harijan[a], Anatoliy Senyshyn[b], R. Ranganathan[c], Dhananjai Pandey[a]

[a]*School of Materials Science and Technology, Indian Institute of Technology (Banaras Hindu University), Varanasi 221005, India.* [b]*Forschungsneutronenquelle Heinze Maier-Leibnitz (FRM-II), Technische Universitat Munchen, Liechtenbergestrass 1, D-85747 Garching b. Munchen, Germany*

[c] *ECMP Division, Saha Institute of Nuclear Physics, 1/AF Bidhannagar, Kolkata 700 064, India*

**Corresponding Author: Dhananjai Pandey,** *School of Materials Science and Technology, Indian Institute of Technology (Banaras Hindu University), Varanasi 221005, India. Ph. No: +91-9453048510, E.Mail: dp.mst1979@gmail.com*

**Abstract:**

The coexistence and coupling of magnetic and ferroelectric orderings in single phase multiferroics has evinced enormous scientific interest as it involves breaking of time reversal and space inversion symmetries in the same material[1]. The mutual controllability of the two diverse orderings in multiferroics has potential for developing new generation multifunctional sensor, actuator and data storage devices [1-3]. We present here evidence for a new exotic facet of multiferroicity, whereby one can raise the strength of antiferromagnetic (AFM) superexchange interaction and hence tune the Néel temperature ($T_N$) from ~120K in bulk to ~350K in 18nm size particles by tuning the ferroelectric distortion in the tetragonal phase of multiferroic $(1-x)BiFeO_3$ -$xPbTiO_3$ solid solutions . This observation is unique to multiferroics only as the $T_N$ in non-multiferroic AFM oxides decreases with particle size[4] . Our results provide a scientific basis for designing room temperature single phase multiferroics, useful for making multifunctional device operating at room temperature.

**Significance:**

Néel transition temperature ($T_N$) in conventional antiferromagnetic (AFM) materials is known to decrease with particle size following the existing theories of critical phenomena in reduced dimension systems. Our results shows that in multiferroic antiferromagnets, the $T_N$ can increase with decreasing particle size because of subtle interplay between the size dependent ferroelectric distortion and the strength of the superexchange interaction. Our work provides a basis for designing room temperature mutiferroics which are very few.

**Introduction :**

Amongst the magnetoelectric mutiferroic compounds, $BiFeO_3$ (BF) is the only room temperature multiferroic with very high ferroelectric (FE) and antiferromagnetic (AFM) transition temperatures of $T_c$ = 1103 K and $T_N$ = 643 K, respectively [1,5]. This material therefore holds considerable promise for multifunctional devices operating at room temperature[1-2, 5]. $BiFeO_3$ shows commensurate ferroelectric and incommensurate magnetic orderings[1]. The structure of BF is rhombohedral in the R3c space group in which the oxygen octahedra in the neighbouring unit cells are antiphase rotated about the trigonal axis[6]. The magnetic ordering in elementary pervoskite unit cell of BF is essentially of G-type antiferromagnetic (AFM) with canted spins[7]. The spin canting angle depends on the angle of antiphase rotation of oxygen octahedra[8]. Because of the canted spins, one expects BF to exhibit weak ferromagnetism[8] and hence linear magnetoelectric coupling[9]. However, bulk BF behaves like a perfect AFM with linear M-H plot[10] as the ferromagnetic component of the canted spins cancel out due to the superposition of a cycloidal incommensurate modulation of the spins with an approximate wavelength of 62 nm[8]. As a result of the AFM nature of BF, symmetry arguments do not allow linear magnetoelectric coupling[9]. The latent weak ferromagnetism of $BiFeO_3$ can be recovered by destroying the spin cycloid using epitaxial constraints[3], application of high magnetic field over ~ 20T[11], size reduction below 62 nm[12] and solid solution formation with other $ABO_3$ perovskites[13]. Among these strategies, studies on solid solutions of $BiFeO_3$ with other perovskites has evoked enormous interest. In this context, studies on the solid solution system $(1-x)BiFeO_3-xPbTiO_3$ (or BF-xPT) have revealed a very interesting phase diagram showing a critical point[14] and a rich variety of structural and magnetic phase transitions like ferroelectric (FE), antiferrodistortive (AFD)[15], ferroelectric to ferroelectric isostructural[14], morphotropic [15,16], antiferromagnetic (AFM)[17], spin reorientation[18] and spin glass (SG) transitions[19]. $PbTiO_3$ substitution modifies the rhombohedral R3c symmetry of BF to monoclinic Cc symmetry[15], similar to that in thin films of BF under epitaxial constraints[20, 21]. The monoclinic Cc phase stable for $x \leq 0.27$ transforms to a tetragonal (space group P4mm) structure for x =0.31 via a thin phase coexistence region for 0.27<x<0.31 across a first order morphotropic phase boundary (MPB)[15, 16]. This letter deals with the magnetic transitions in a tetragonal composition of BF-xPT with x = 0.50 (BF-0.50PT).

The tetragonal phase of BF-xPT stable for x = 0.31 has got several unusual characteristics : (1) it shows anomalously high tetragonality ($\eta$= (c/a -1)) that is more than three times that of the commercial MPB ceramics like PZT, (2) its tetragonality increases on approaching the MPB and becomes ~ 18.73% for x=0.31[15,16] whereas it decreases on approaching the MPB in PZT,(3) it does not result from the paraelectric (PE) cubic phase directly but by an isostructural phase transition from a low $\eta$ phase, which gets formed first from the cubic paraelectric phase below the Curie temperature[14], (4) it can also result from a stress induced phase transition from the mononclinic Cc phase [21] and (5) it shows a very low $T_N$ ( ~ 209K for x=0.31) as compared to neighbouring monoclinic compositions with $T_N$ ~ 473 K for x=0.27[17]. There is no satisfactory explanation for this drastic drop in $T_N$ across the MPB. Here, we present evidence for extraordinary tunablity of the $T_N$, in such tetragonal compositions taking BF-xPT powders with x= 0.50 (i.e., BF-0.5PT) of different sizes such that $T_N$ increases from ~120K in bulk to ~350K in 18nm size powders .

**Results and Discussion:**

BF-0.50PT powders of approximate average sizes of 120nm, 45nm, 31nm and 18nm were synthesized by the sol-gel route, the details of which are given in the supplementary information (section A). Figure 1 shows the Synchrotron x-ray powder diffraction (SXRPD) profiles of the 100, 110 and 111 pseudocubic reflections of BF-0.5PT with average particle size of ~ 18, 31, 45 and 120 nm. The SXRPD profiles of 120 nm size particles show that the 100 and 110 peaks are doublet while 111 is a singlet as expected for the tetragonal phase with P4mm space group. With decreasing particle size, the splitting of 100 and 110 peaks decreases and the profile broadening increases. There is an asymmetric broadening of the 001 peak and other peaks with decreasing size. The 120 nm powder possesses a single tetragonal phase (space group P4mm) whereas the smaller size powders show coexistence of two tetragonal phases, a dominant tetragonal phase with higher tetragonality and a minority phase with relatively lower tetragonality, both in the P4mm space group, as confirmed by Rietveld refinement using Synchrotron x-ray powder diffraction (SXRD) data. The details of Rietveld refinements are presented in the supplementary information (section B). The 18 nm size particles do not show any obvious splitting of 200 and 220 and may appear to be cubic in the first appearance. However Rietveld analysis of the SXRD data (as explained in supplementary section B) shows that the sample is still tetragonal with a tetragonality of 5.75%. The spontaneous polarisation of the dominant tetragonal phase of BF-0.5PT as obtained using Born effective charges of various cations and anions in BF[22] and PT[23] and the refined positional coordinates of BF-0.5PT are 32.1, 28.2, 24 and 23.1 $\mu C/cm^2$ for 120, 45, 31 and 18nm size particles, respectively, confirming that significant ferroelectric polarization is present at room temperature even in the 18nm particles. The reduction in tetragonality and spontaneous polarization with size is due to well known 'size effect' in ferroelectics caused by the competition between depolarization and domain wall energies[24].

Fig. 2 (a to d) depicts the results of magnetization (M) measurements as a function of temperature (T) at an applied field of 5000 Oe for 120 nm, 45 nm, 31 nm and 18 nm size powders in zero-field cooled (ZFC) and field cooled (FC) modes. The ZFC M-T plot of the 120 nm size particles shows a Néel transition temperature $T_N \sim 120K$ (see the inset also) similar to that reported in bulk powders of BF-0.5PT[19]. Intriguingly on reducing the particle size to 45 nm, 31 and 18 nm, the Néel temperature increases drastically to 230, 250 and 350K, respectively, as can be seen from various insets to Fig 2. The variation of Néel temperature with size has been shown in the Fig 2 (f).

The enhancement of the Néel temperature from $T_N \sim 120K$ in the bulk powders to $T_N \sim 350K$ in 18 nm size powders of BF-0.50PT was confirmed by the disappearance of the magnetic peak corresponding to propagation vector **k**=½ ½ ½ in the neutron powder diffraction patterns shown in Figs 3(a) and 3(b). The variation of the integrated intensity of the ½ ½ ½ magnetic peak with temperature shown in Fig 3 (c) reveals $T_N \sim 120K$ and $\sim 350K$ at which the integrated intensity of the magnetic peak becomes zero for the bulk (120nm) and 18nm size powders, respectively, in agreement with the $T_N$ determined from ZFC M-T measurements (Fig. 2 (a) and (d)).

The enhancement of $T_N$ with size reduction is unique to multiferroics, since $T_N$ in non-multiferroic AFM oxides is known to decrease with particle size [4, 25-26]. Finite size effects in non-multiferroic AFMs like NiO and CoO nanoparticles and CoO and Holmium thin film layers have been investigated very systematically [25, 27-30] and it has been shown that $T_N$ decreases with decreasing particle size or the thin film thickness. It has also been reported [25] that the $T_N$ shows a power law dependence on grain size in agreement with Binder's theory of critical phenomena in reduced dimension systems [31] as per the following relationship:

$$\frac{T_N^b - T_N^{grain}}{T_N^b} = \left(\frac{d}{\xi_0}\right)^{-(1/\nu)}$$

where $T_N^b$ and $T_N^{grain}$ are the bulk and finite size AFM transition temperature, d is the grain diameter, $\xi_0$ is the magnetic correlation length at 0K and ν the critical exponent related to ξ. In the case of BF-0.5PT, the situation is just opposite to the conventional AFM system as $T_N$ increases with decreasing particle size.

The $T_N$ enhancement in BF-0.50PT with size reduction is a consequence of the subtle interplay between the ferroelectric distortion and the strength of the superexchange interaction in the multiferroic BF-0.5PT. The longer $Fe^{3+}$- $O^{2-}$ bond length in the [001] direction (labelled as Fe-$O_{1b}$ in Table I) in the bulk powder is too big (2.571 Å) as compared to the sum of the ionic radii of $Fe^{3+}$ and $O^{2-}$ (~2.045 Å) to have any orbital overlap required for superexchange interactions. As a result, the AFM superexchange interaction pathways are confined to [100] and [010] directions only. Since the longer $Fe^{3+}$-$O^{2-}$ bond in the [001] direction decreases systematically with particle size reduction (see Table I), it leads to gradual enhancement of superexchange interaction pathways in the [001] direction also. This crossover from essentially 2d AFM interactions in bulk to 3d AFM interactions in nanoparticles of BF-0.50PT is responsible for the drastic rise in the $T_N$ as a result of size reduction. Thus, the physics of size effect in multiferroic AFM systems is entirely different from that in conventional AFM systems.

The magnetic neutron peaks of 18 nm particles have sharp and diffuse (broad) components (see Fig 3(d)) linked with the coexisting high and low tetragonality phases, respectively, confirmed by Rietveld refinements using SXRD and neutron powder diffraction data, the details of which are given in supplementary file (sections B and C). The diffuse component of the magnetic scattering is dominant above 250K and shows a temperature dependence different from that of the sharp component. We propose that the diffuse component of the magnetic peak with shorter correlation length is due to spin disorder in a thin shell region of the surface layers of the nanoparticles. In this scenario, one expects weak ferromagnetism due to spin canting in the shell region of the 18 nm size nanoparticles of BF-0.5PT as observed in similar core-shell type systems in non multiferroic AFM oxides[32]. This weak ferromagnetism should be present even at room temperature as the corresponding $T_N$ is well above room temperature for 18 nm size particles. This was confirmed by the M-H measurements, the results of which are shown in Fig 4. It is evident from Fig 4(a) that the M-H plots for 120 nm, 45 nm and 31 nm size powders are straight lines confirming their paramagnetic nature at room temperature, whereas the 18 nm size particles exhibit nonlinear M-H curve, with slight loop openings. With decreasing temperature, the M-H loop of 18 nm size particles opens up further as can be seen from the two insets of Fig 4(b) which compare the M-H plots for 18 nm size particles at 300 K and 5 K. Further, the M-H loop in Fig 4(b) inset reveals an exchange bias ($E_B$) of ~ 337 Oe at 5K, which decreases with increasing temperature for the same particle size. The $E_B$ decreases with increasing particle size also for the same temperature. For example, the $E_B$ at 5 K is ~17 Oe and ~8 Oe for 31 nm and 45 nm size powders, respectively, while it is absent for the bulk powder (see Fig 4(c) inset). The more pronounced $E_B$, as shown in the inset of Fig 4(b), in the smallest size particles (~18 nm) is due to higher surface to volume ratio of 18 nm particles leading to more number of spins in the surface layers constituting the shell region. The interface between the FM component of the canted spins in the shell region and the collinear AFM spins in the core is responsible for the exchange bias effect as observed in other systems also [32-34]. Thus our M-H plots showing $E_B$ effect not only confirm the coexistence of AFM and weak FM ordering but also reveals that they arise from a core-shell type of nanostructure of individual nanoparticles.

Before we close, we would like to discuss briefly the origin of the cusp at $T_f$ ~17K in the ZFC M-T plot of the bulk sample which has been attributed to spin glass freezing with opening of a slim M-H loop below $T_f$. The cusp is suppressed in the field-cooled M-T data for the bulk samples (Fig 2 (a)). For the smaller size particles, the low temperature anomaly is not seen in ZFC M-T plots (see Fig. 1 b to d), until the field is lowered to ~ 500 Oe as shown in Fig 2(e) for 18nm size powders. Even for 500 Oe field, the low temperature anomaly is very broad and the $T_f$ is enhanced to ~50K (see Fig. 2(e)). The broad anomaly in Fig 2(e) does not seem to be due to spin glass freezing, as we did not observe any measurable shift in $T_f$ in the temperature dependent AC $\chi$ measurements as a function of frequency. The nature of M(T) data seems to suggest that this broad anomaly may be due to a superparamagnetic (SPM) blocking[35] over a range of temperatures with a bifurcation temperature $T_{bif}$ ~320 K between ZFC and FC M(T) plots .

To summarise, we have shown that size reduction in the tetragonal phase of $(1-x)BiFeO_3-xPbTiO_3$ can be used to raise the AFM Néel transition temperature ($T_N$) drastically from 120K to 350K for x=0.50 by increasing the strength of the superexchange interactions via a decrease in the ferroelectric distortion in nanocrystalline powders. This enhancement of $T_N$ with decreasing particle size is unique to multiferroics as the $T_N$ in conventional AFM oxides like CoO and NiO decreases with particle size as per the theory of critical phenomena in reduced dimension system. We have also shown that size reduction can induce weak ferromagnetism and exchange bias due to AFM core and weakly FM shell type magnetic nanostructure. We believe that our results will stimulate more work as it has tremendous potential for designing room temperature multiferroics by tuning the $T_N$ to room temperature through a manipulation of the ferroelectric distortion via size reduction, a phenomenon useful for multifunctional devices operating at room temperature.

**Method:** Samples were synthesized by sol-gel route as explained in the Section A of Suppleimentary information. For the optimization of the calcination and sintering conditions, phase identification was carried out by X-ray diffraction (XRD) measurements using an 18 kW rotating anode (CuKα) based Rigaku (RINT 2000/ PC series) powder diffractometer operating in the Bragg- Brentano geometry and fitted with a curved graphite crystal monochromator in the diffracted beam. The operating voltage and current were kept at 40 kV and 150 mA, respectively.

Average particle size was estimated by Scanning Electron Microscopy using a Carl-Zeiss Supra 40, high resolution, field emission gun scanning electron microscope (FEG SEM) under optimized operating voltage.

Synchrotron X-ray diffraction patterns were recorded at high resolution powder diffraction beamline P02.2 at PETRAIII, DESY, Hamburg, Germany operating at 60keV X-rays. The wavelength for the synchrotron x-ray used was 0.2079 Å.

Magnetic characterization was carried out using a Quantum Design® Evercool II PPMS®. Magnetization vs temperature measurements were performed in the temperature range of 2.5K-390K in an applied field of 5000 Oe or 500 Oe on case to case basis, whereas magnetization vs applied field measurements were performed at selected temperatures up to a maximum field of 7 Tesla.

Neutron powder diffraction measurements were carried out on high-resolution powder diffractometer SPODI at FRM-II, Garching, Germany. The incident neutron wavelength was 2.5367 Å. Approximately 10 g of the sample, contained in a cylindrical niobium holder with 50 $\mu$m wall thickness, 40 mm height, and 10 mm diameter, was used for these measurements. The data were collected at steps of 0.05° in the 2*θ* range from 1 to 157°. A closed cycle cryostat was used for sample temperature variation in the temperature range of 4 K to 500 K

**Acknowledgements:** D. Pandey acknowledges financial support from Science and Engineering Research Board (SERB) of India through J.C. Bose Fellowship grant. P.K. Harijan acknowledges financial support under Rajiv Gandhi National Fellowship (RGNF) programme of India. We acknowledge the assistance of Dr. M. Hinterstein from PETRA III, DESY, Germany in collection of SXRD data. We also acknowledge financial support from Saha Institute of Nuclear Physics (SINP) under the DST-DESY project to carry out SXRD experiments at DESY, Germany. We thank Anar Singh for useful discussions.

**Figure caption**

Fig.1.: **Synchrotron powder diffraction analysis of BF-0.50PT with size.** SXRPD profiles of 100, 110 and 111 pseudocubic reflections of the samples with average particle size of ~ 18, 31, 45 and 120 nm. The SXRPD profiles of 120 nm size particles show that the 100 and 110 peaks are doublet which started merging with the decrease of particle size.

Fig.2. (a-d): **The FC and ZFC magnetization vs temperature for different nanoparticles of BF-0.5PT**. The temperature-dependence of the FC and ZFC magnetizations of BF-0.5PT powders of 120, 45, 31 and 18nm sizes under magnetic field of 5000 Oe. The inset shows magnified view of the plot near Neel temperature ($T_N$). Note the enhancement in the $T_N$ with decreasing particle size. (e) Temperature variation of ZFC and FC magnetization under 500 Oe for 18 nm size powders. The inset of (e) shows the difference in FC and ZFC magnetization value. (f) Variation of Néel Temperature with size.

Fig.3. **Neutron powder diffraction analysis of BF-0.50PT with temperature.** The evolution of the magnetic peak corresponding to k= 1∕2 1∕2 1∕2 with temperature for (a) 120 and (b) 18 nm size powders of BF-0.5PT (c) Variation in the integrated intensity of the magnetic peak with temperature for 120 and 18 nm size powders. Dots represent measured data points, while the continuous line corresponds to the least-squares fit. (d) Deconvolution of the magnetic peak profile (k=1∕2 1∕2 1∕2)$_{pc}$ of 18 nm size powder showing different magnetic correlation lengths (ξ).

Fig.4. **Magnetization (M) vs field hysteresis of different particle size samples of BF-0.50PT**. (a) M-H loops for 18, 31, 45,120 nm size powders at 300K. The inset gives a magnified view to show the linear M-H plot for all the powders, except 18 nm for which M-H plot is non-linear. (b) M-H loops for 18 nm particles at 300 and 5K . The two insets show exchange bias which is much higher at 5K. (c) M-H loops for 18, 31, 45, 120nm size powders at 5K.

**Table I:** Fe-O bond lengths (Å) and tetragonality of the main phase in BF-0.5PT samples of different sizes for the dominant tetragonal phase.

| Size (nm) | Tetragonality ($\eta$) % | Bond length (Å) | | |
|---|---|---|---|---|
| | | Fe-O$_{1a}$ | Fe-O$_{1b}$ | Fe-O$_2$ |
| 18 | 5.75 | 1.787 | 2.348 | 2.013 |
| 31 | 6.64 | 1.790 | 2.369 | 1.970 |
| 45 | 9.60 | 1.792 | 2.460 | 1.973 |
| 120 | 14.16 | 1.822 | 2.571 | 1.968 |

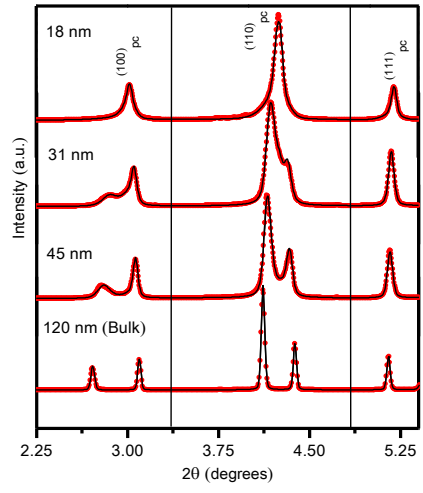

Fig. 1

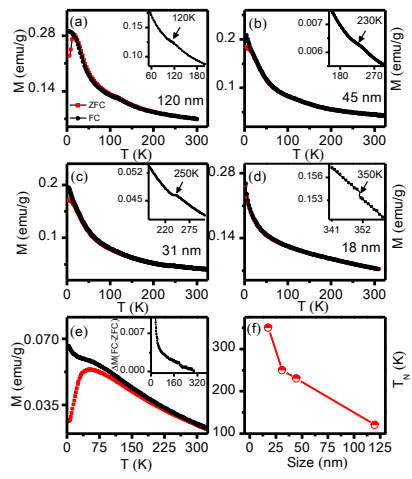

Fig. 2

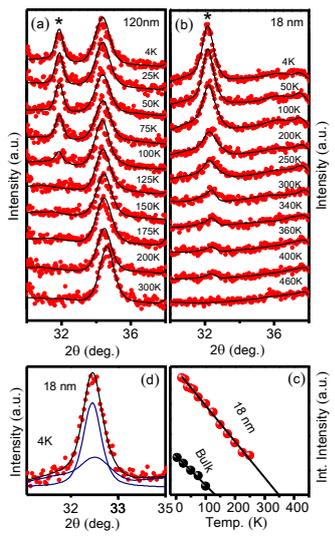

Fig. 3

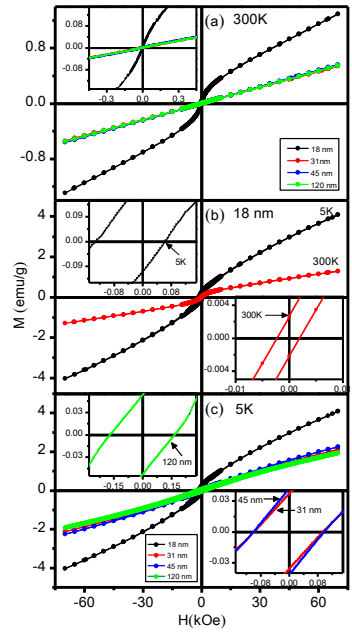

Fig. 4

# Extraordinary Tunability of the Superexchange Interactions in Nanoparticles of the Multiferroic 0.50BiFeO$_3$-0.50PbTiO$_3$

## A. Synthesis of the samples

Powders of four different sizes of (1-x)BiFeO$_3$-xPbTiO$_3$ solid solutions with x=0.50 (i.e., BF-0.5PT) powders were synthesized by sol-gel route. Stoichiometric amounts of metal nitrate (iron, lead and bismuth) aqueous solutions and ethylene glycol (EG) stabilized titanium isopropoxide solution were mixed together at 60°C. In order to avoid any precipitation of TiO$_2$ by moisture or air (i) EG has been taken 7 times by volume as compared to the volume of titanium isopropoxide solution and (ii) the pH was adjusted between 1 and 2. This mixed solution was added slowly to the aqueous solution of maleic acid kept at 60°C and the resultant solution was stirred continuously to ensure complexation of metal cations. The solution was then heated to 80°C to initiate evaporation of water and polymerization. The overnight heating of the sample resulted in a brown coloured gel. The excess ethylene glycol present in the gel was dried at 150-170°C, leaving behind a dark-brown dried polymer precursor. The precursor was then calcined at 550 °C for 6 hours, in a closed environment. The powder so obtained has an average particle size of 18 nm.

The as calcined powder was crushed into fine powder using an agate mortar and pestle. 2% polyvinyl alcohol (PVA) solution was used as binder. A cylindrical steel die of 12 mm diameter was used to make green pellets of BF-0.5PT powder. The die containing the powder was uniaxially pressed at an optimised pressure of 65 kN in a hydraulic press. The green pellets were kept on an alumina plate and heat treated in a furnace at 500°C for 10 hours to burn off the binder material (PVA). Sintering of the PVA evaporated pellets was carried out in a closed alumina crucible sealed with MgO powder in the temperature range of 550°C to 1050°C for 6 hours. $Bi_2O_3$ atmosphere was maintained inside the closed alumina crucible by keeping suitable amount of the as-calcined powder of the same composition inside the closed alumina crucible as a spacer powder. This sintering in closed environment at different temperatures for 6 hours in different batches resulted into samples with varying sizes as indicated in Table SI. The sintered pellets so obtained were crushed into fine powders and then annealed at 500°C for 10 hours to remove the strain introduced during crushing. Such annealed powders were used for diffraction studies.

Scanning electron micrographs were used to estimate the average particle size of the powders. The SEM images of the powders used in this investigation are given in Fig. S1. The average size and sintering temperatures are given in Table S1. The minimum particle size has been found to be ~18 nm whereas the largest size is ~120 nm corresponding to the sintering temperatures of 550 and 1050 $^0$C, respectively.

**Table SI:** Variation of particle size of the BF-0.50PT samples with sintering temperature.

| Sintering Temperature ($^0$C) | Average particle size (nm) |
|---|---|
| 550 | 18 |
| 850 | 31 |
| 950 | 45 |
| 1050 | 120 |

**B. Supplemental details of structure refinement using synchrotron x-ray powder diffraction data.**

Synchrotron x-ray powder diffraction (SXRPD) patterns were used for Rietveld refinements carried out with FULLPROF package[1]. Pseudo-Voigt function was used to model the profile shape. Anisotropic peak broadening was modelled using Stephen's model[2]. The background was modelled using linear interpolation method.

In Fig. 1 of the main text, the 18 nm size particles do not show any obvious splitting of 200 and 220 peaks and therefore the structure may appear to be cubic in the first appearance. However, the Rietveld fit with single phase cubic Pm3m space group leads to rather poor fit with very high $\chi^2$ value (101.9) (see Fig.S2a of the supplementary file ). The asymmetric broadening towards lower $2\theta$ angle of the 200 and 220 pseudocubic peaks suggests that the structure may have lower symmetry than cubic. Accordingly, we considered the tetragonal space group P4mm for refinement. While this led to considerable improvement in $\chi^2$ (from 101.9 to 62.41), the $\chi^2$ value is still large (Fig. S2b). At this stage, we considered Rietveld refinement using coexistence of tetragonal and cubic phases which led to further improvement in $\chi^2$ (from 62.41 to 34.24) but still the fits are not good (Fig.S2c). Finally, we attempted to fit the diffraction data with two tetragonal phases with different tetragonalities but the same space group (P4mm). This led to the best possible fit shown in Fig S2-d. with significantly improved $\chi^2$ value (15.70). This indicates that the samples having average particle size of 18 nm consist of two coexisting tetragonal phases with different tetragonalities. As explained later on in the analysis of the powder neutron diffraction patterns that this is due to a core-shell type nanostructure with different tetragonalities in the core and shell regions of each particle. Rietveld refinements for the different size powders confirmed that the 120 nm powder consists of a single tetragonal phase, whereas all the smaller size powders contain two coexisting tetragonal phases. Fig. S3 depicts the Rietveld fits for the remaining samples. The c/a ratio of the dominant tetragonal phase decreases with decreasing particle size as show in Fig. S4. This is expected as per the well known size effects in ferroelectric perovskites [3,45].

**C. Supplemental details on magnetic structure refinements**

The Rietveld refinement of the magnetic structure of BF-0.5PT was carried using the neutron powder diffraction data on 18 and 120 nm size powders using representation theory. The irreducible representation (Ireps) for tetragonal P4mm and cubic Pm3m space group were used in the refinements [6,7]: As in case of Rietveld refinement of the nuclear structure using SXRPD patterns, we considered cubic Pm3m, tetragonal P4mm, coexistence of cubic and tetragonal and coexistence of two tetragonal phases in the simultaneous refinements of the nuclear and magnetic structures. Attempts to fit single cubic Pm3m phase, single tetragonal P4mm phase and coexistence of cubic and tetragonal phases led to unsatisfactory fits and significantly larger $\chi^2$ values in agreement with similar findings discussed in the supplementary information on SXRPD studies (section C). The best fit with the lowest $\chi^2$ values was obtained for the coexistence model consisting of two tetragonal phases, both in the P4mm space group, but with different tetragonalities as illustrated in the Figure S5. Table SII lists the refined parameters obtained from simultaneous refinement of the nuclear and magnetic structures of 18 nm size BF-0.5PT powders. For the bulk powder, the refined parameters are comparable to those reported in the literature[6].

**Table SII:** Rietveld refined positional coordinates, thermal parameters and lattice parameters of 18 nm size powder of BF-0.50PT at 4K. The magnetic R-factor for the coexisting phase is rather high due to the diffuse nature of the neutron peak component.

| BF0.5PT with different sizes | | Fractional coordinates | | | | Thermal Parameters (Å$^2$) | Lattice Parameters (Å) | Statistical parameters |
|---|---|---|---|---|---|---|---|---|
| | | Atom | X | y | z | | | |
| 18 nm | Tetragonal phase-1 | Bi/Pb | 0.0 | 0.0 | 0.0 | $B_{11}=B_{22}$= 0.146(4), $B_{33}$=0.183(9) | a=b= 3.9075(4) c= 4.129(1) | $R_p$= 10.0 $R_{wp}$= 9.49 $R_{exp}$= 7.37 $\chi^2$ = 1.659 Magnetic R-factor: 1.27 (sharp component) Magnetic R-factor: 17.8 (diffuse component) |
| | | Fe/Ti | 0.5 | 0.5 | 0.509(3) | $B_{iso}$=2.4(2) | | |
| | | O1 | 0.5 | 0.5 | 0.121(2) | $B_{11}=B_{22}$= 0.019(6) $B_{33}$= 0.05(1) | | |
| | | O2 | 0.0 | 0.5 | 0.649(1) | $B_{11}=B_{22}$= 0.091(5) $B_{33}$= 0.070(6) | | |
| | Tetragonal phase-2 | Bi/Pb | 0.0 | 0.0 | 0.0 | $B_{11}=B_{22}$= 0.028(2), $B_{33}$=0.005(3) | a=b= 3.9458(2) c= 4.0026(5) | |
| | | Fe/Ti | 0.5 | 0.5 | 0.599(2) | $B_{iso}$=2.4(1) | | |
| | | O1 | 0.5 | 0.5 | 0.056(2) | $B_{11}=B_{22}$= 0.052(2), $B_{33}$= 0.00006(1) | | |
| | | O2 | 0.0 | 0.5 | 0.576(1) | $B_{11}=B_{22}$= 0.005(3), $B_{33}$= 0.140(4) | | |

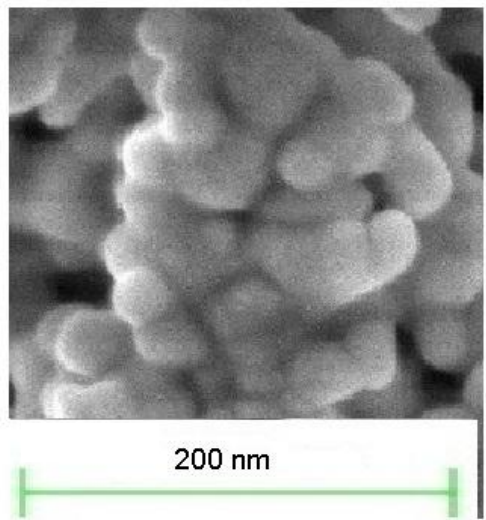
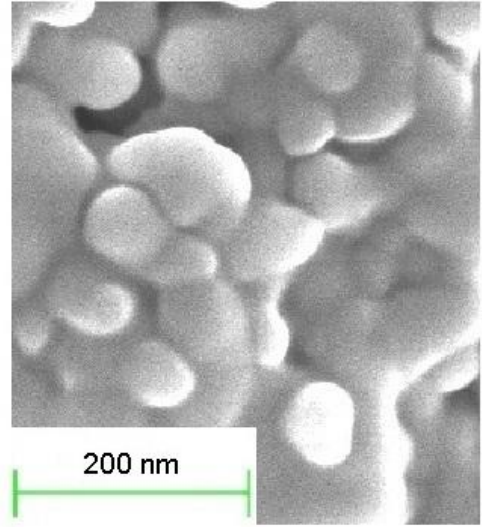
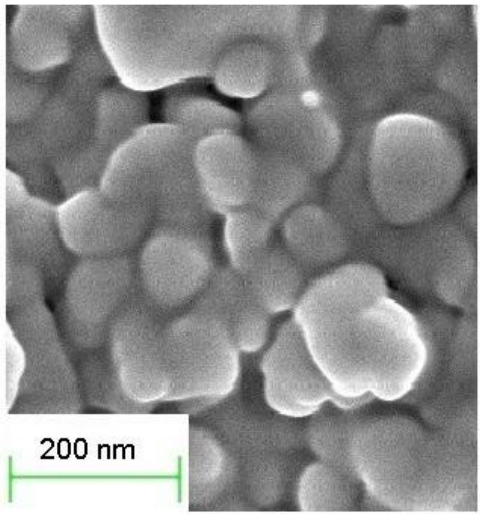
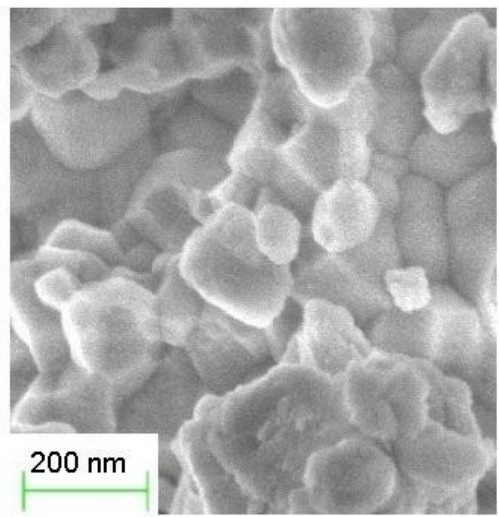

(a)  (b)  (c)  (d)

Fig. S1: Scanning electron micrographs of BF-0.50 PT samples sintered at different temperatures (a) 550 $^0$C, (b) 850 $^0$C, (c) 950 $^0$C and (d) 1050 $^0$C

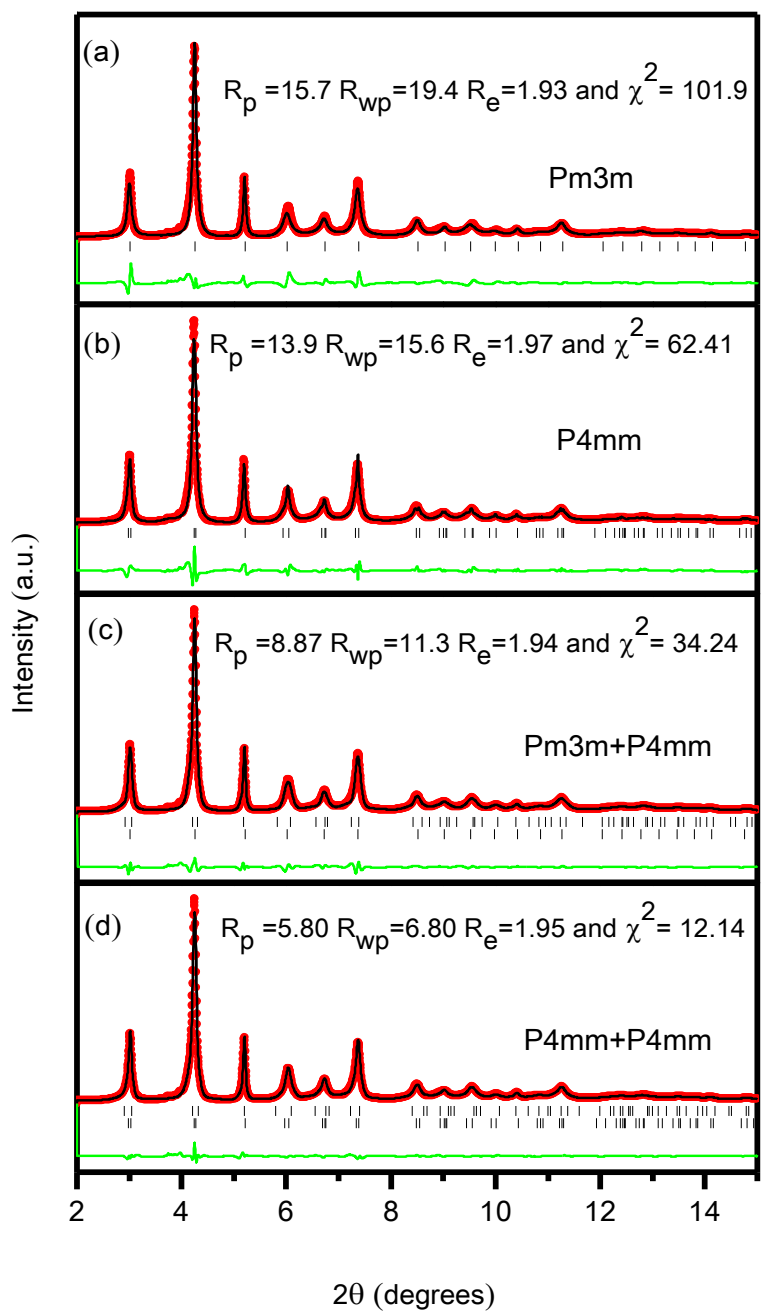

Fig S2: Rietveld fit of synchrotron X-ray powder diffraction pattern of BF-0.50PT powder of 18 nm particle size using different structural models.

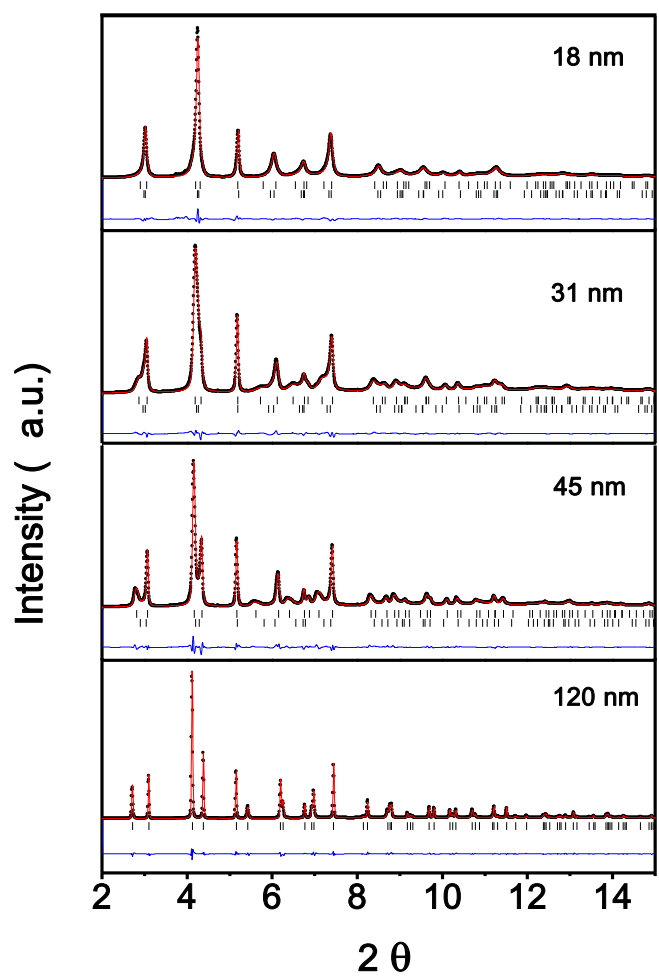

Fig S3: Rietveld fit of synchrotron X-ray powder diffraction pattern of BF-0.50PT powder of different particle size using two tetragonal phases.

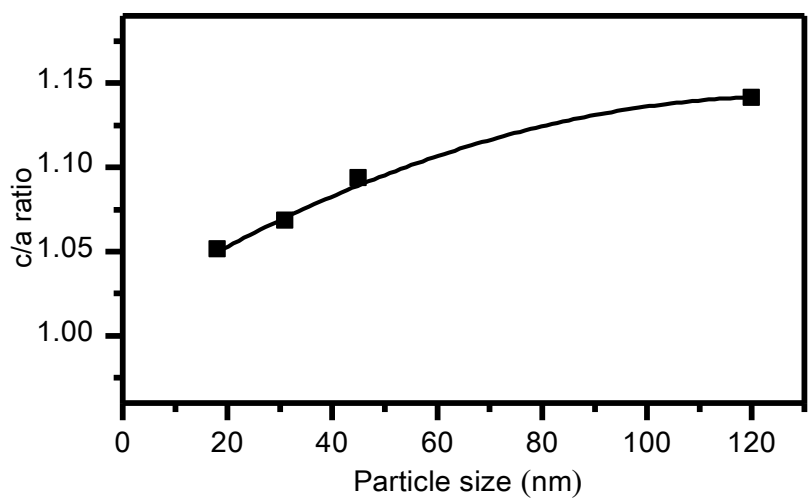

Fig S4: Variation of the c/a ratio of core with particle size. The error bars within the size of the data points.

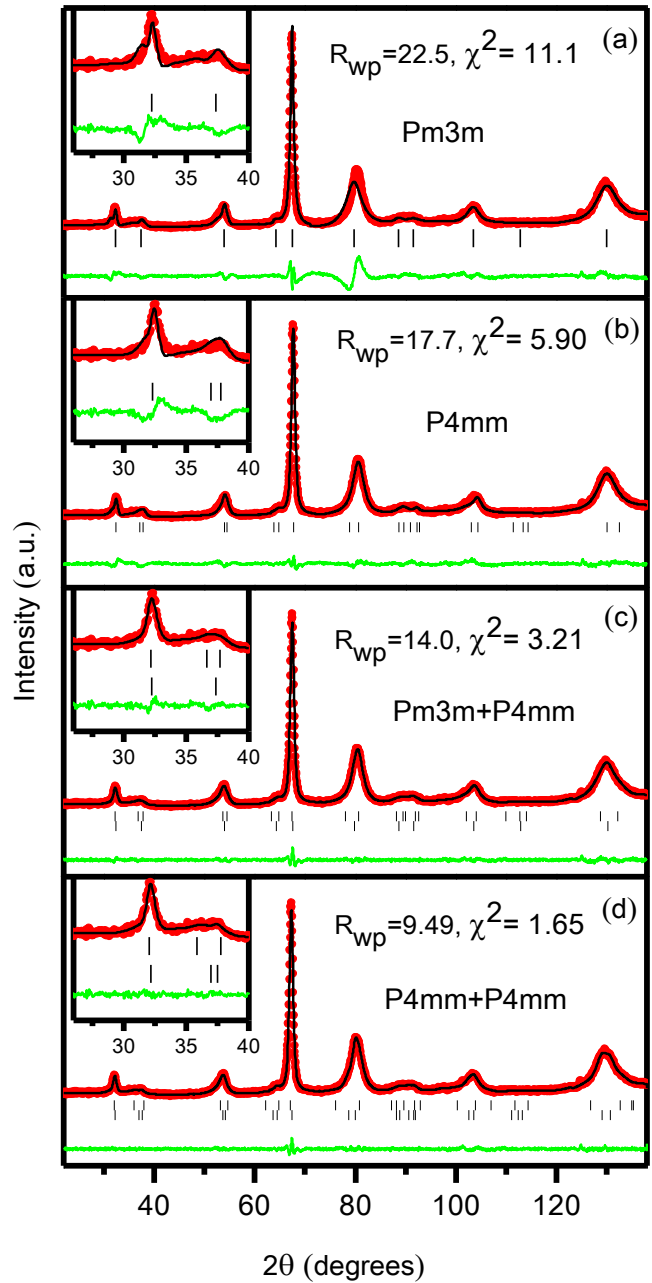

Fig.S5: Rietveld fit of neutron powder diffraction patterns of BF-0.5PT powder recorded at 4 K for 18 nm particle size at 4K using different structural models.